\documentclass{mem}
\usepackage{natbib}\usepackage{txfonts}\usepackage{balance}
\usepackage{graphicx}
\usepackage[a4paper]{hyperref}
\idline{75}{282}
\begin{document}
\def\teff{$T\rm_{eff }$}
\def\kms{$\mathrm {km s}^{-1}$}


\newcommand{\dss}{$\delta$~Scuti}
\newcommand{\epscep}{$\epsilon$~Cep}
\newcommand{\epscepp}{$\epsilon$~Cephei}
\newcommand{\lamsco}{$\lambda$~Sco}
\newcommand{\lamscoo}{$\lambda$~Scorpii}
\newcommand{\str}{Str\"omgren}
\newcommand{\betacep}{$\beta$ Cep}
\newcommand{\betacepp}{$\beta$~Cepheid}
\newcommand{\eg}{e.g.}
\newcommand{\ie}{i.e.}
\newcommand{\cf}{c.f.}
\newcommand{\lee}{\emph{left}}
\newcommand{\rii}{\emph{right}}

\title{
High-precision photometry with the WIRE satellite
}

   \subtitle{}

\author{
H.\ Bruntt\inst{1} 
\and D.\ L.\ Buzasi\inst{2}
          }

  \offprints{H. Bruntt}

\institute{
Niels Bohr Institute,
Juliane Maries Vej 30,
DK-2100 Copenhagen \O,
Denmark
\email{bruntt@phys.au.dk}
\and
US Air Force Academy, CO, USA
\email{derek.buzasi@usafa.af.mil}
}

\authorrunning{Bruntt}

\titlerunning{High-precision photometry with the WIRE satellite}

\abstract{Around 200 bright stars ($V<6$) have been monitored
with the two-inch star tracker on the WIRE satellite since observations 
started in 1999. Here we present new results for the solar-like star Procyon~A, 
the two \dss\ stars Altair and \epscepp, and the triple system \lamscoo\ which consist 
of two B-type stars -- one of which we find to be an eclipsing binary.
\keywords{Stars: variable, Stars: individual: Procyon~A, Altair, \epscepp, \lamscoo}
}
\maketitle{}

%
\begin{figure}[]
\resizebox{\hsize}{!}{\includegraphics[clip=true]{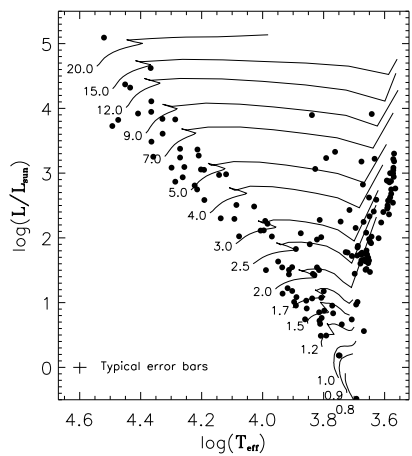}}
\caption{\footnotesize
HR diagram showing the WIRE targets.
}
\label{fig:hr}
\end{figure}
%


\begin{figure}[]
\resizebox{\hsize}{!}{\includegraphics[clip=true]{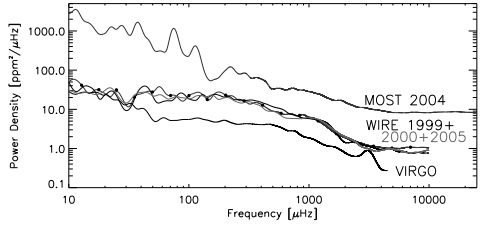}}
\caption{
\footnotesize
Power density spectrum for VIRGO measurements of the Sun and
data from the WIRE and MOST satellites for Procyon~A.
}
\label{fig:procyon2}
\end{figure}

\section{Introduction}

The Wide-Field Infrared Explorer 
(WIRE) was launched in 
1999 and designed to make a survey 
of star-burst galaxies \citep{hacking99}.
Unfortunately, the coolant for the main camera leaked and the
main mission failed.
However, since then the star tracker has been used as the 
first asteroseismology instrument 
in space (\cite{buzasi00}). Oscillations have been detected in 
the solar-like star $\alpha$~Cen~A (\cite{schou}) and
the K~giant Arcturus (\cite{retter}), while Altair
has been found to be a low-amplitude $\delta$~Scuti star (\cite{buzasi05, suarez05}).

About 200 targets have been observed by WIRE from 1999--2005 and their
location in the Hertzsprung-Russell (HR) diagram are shown in Fig.~\ref{fig:hr}.  
Temperatures were found using Str\"omgren indices with 
TEMPLOGG \citep{rogers95} and we used HIPPARCOS parallaxes 
and bolometric corrections from \cite{bessell98} to estimate luminosities.
In the lower left corner typical error bars on the
luminosity and temperature are shown.
Also plotted are evolution tracks for solar metallicity from \citet{lejeune}; 
masses in solar units are indicated. 
Each star has been monitored continuously for typically 2--4 weeks.

In the following we will present some recent results obtained in 2005
for stars in different parts of the HR diagram. In order of increasing mass,
we show results for the solar-like star Procyon~A, 
the two \dss\ stars Altair and \epscep,
and the \betacepp\ type variable
\lamsco, which is an eclipsing triple system.

\section{A Solar-like star: Procyon~A}

Procyon~A was monitored for 9.5 days in
September/October 1999, 7.9 days in September 2000,
and 19.7 days in March/April 2005. The first
two datasets were analysed by \citet{bruntt05}.
The duty cycle is 18\% in the datasets from
1999 and 2000 and 30\% in the dataset from 2005. 
As a result of the low duty cycle there is 
significant power in the amplitude spectrum at 
the harmonics of the orbital frequency of around 174\,$\mu$Hz.
This can be minimized by filtering out peaks in the low frequency
part of the spectrum which leak power at the harmonics of the orbital frequency.

In Fig.~\ref{fig:procyon2} we show the power density spectrum (PDS)
for Procyon~A for three light curves from WIRE after some high-pass
filtering (see \citet{bruntt05} for details) and we compare the PDS
with results from the Sun and observations of Procyon~A from MOST
\citep{matthews04}. The much higher noise level seen in the MOST
data could be due to scattered light (see \citet{bedding05}). 
The increase of the noise seen in the PDS from WIRE 
when going towards lower frequencies and the plateau
below 300$\mu$Hz is similar to that seen in the Sun. 
We have interpreted the signal seen in Procyon~A 
as being due to the granulation.
In the range 100--300$\mu$Hz the granulation {\em amplitude} 
is about 1.8 times higher in Procyon~A compared to the Sun, 
which is expected since granulation in Procyon~A should 
be more violent since it is both hotter and more evolved.

%

\begin{figure}[]
\resizebox{\hsize}{!}{\includegraphics[clip=true]{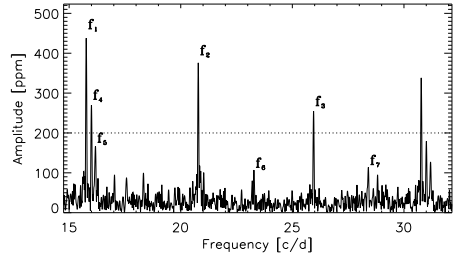}}
\caption{
\footnotesize
Amplitude spectrum of Altair from WIRE. 
}
\label{fig:altair}
\end{figure}

\begin{figure*}[]
\resizebox{\hsize}{!}{\includegraphics[clip=true]{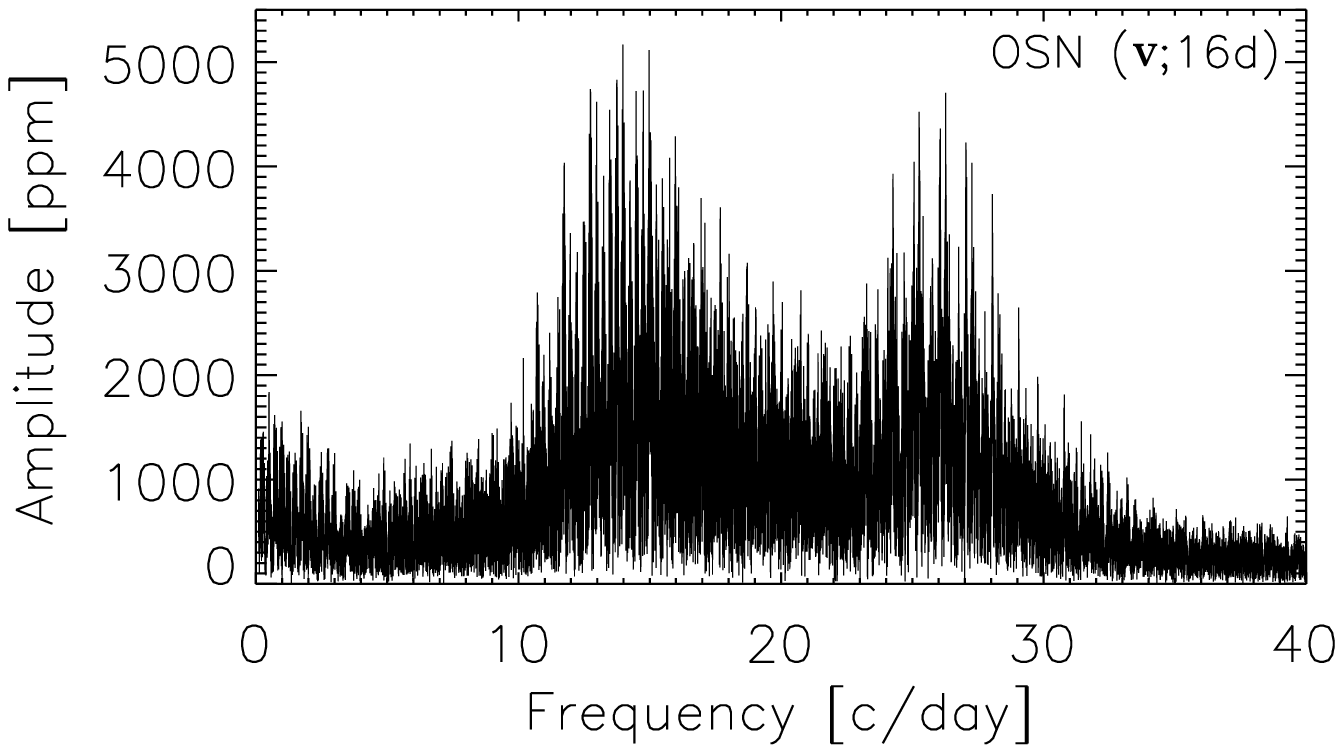}
                      \includegraphics[clip=true]{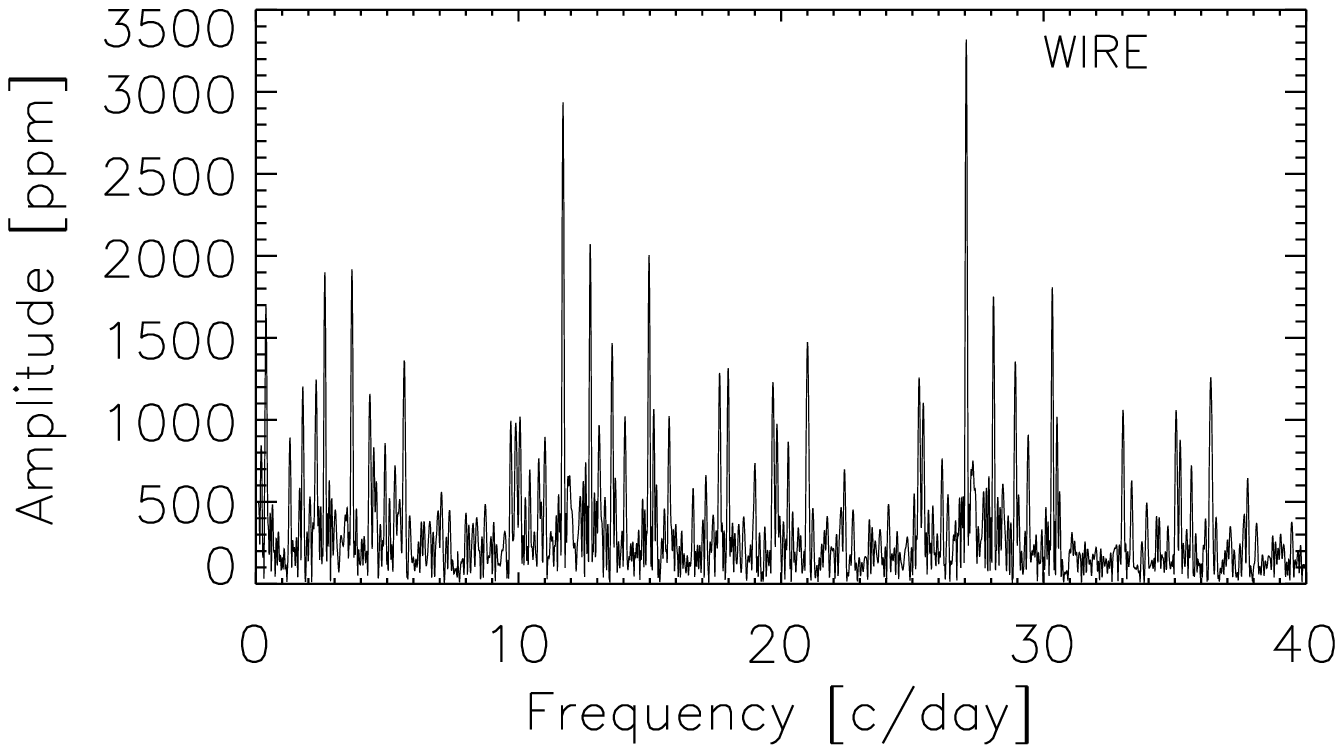}}
\caption{
\footnotesize
Amplitude spectra of the \dss\ star \epscep\ from OSN and WIRE.
}
\label{fig:epscep}
\end{figure*}

\section{\dss\ stars: Altair and \epscep}

Altair was observed for about 22 days in 
October/November 1999 and \citet{buzasi05} 
reported that the star is a multi-periodic \dss\ star with at least 
seven excited modes. Even though the star is the 12th brightest star 
in sky this was not known previously, since it is too 
bright ($V=0.8$) to observe for even small ground-based telescopes.

The amplitude spectrum of Altair is shown in Fig.~\ref{fig:altair} where
the seven modes are indicated. Note the peaks above 30 c/day which
are aliases due to the combinations $[f_1-f_3] + f_W$, where $f_W$ is the
orbital frequency of WIRE which is 15.0 c/day.

\citet{suarez05} tried to model Altair but found that the rapid
rotation of the star makes this difficult. According to
\citet{royer02} Altair has $v \sin i > 200$\,\kms. 


We are currently planing to observe a number of moderately rotating
\dss\ stars with WIRE and from ground-based observatories with multi-colour
photometry. Our goal is to identify the modes and 
to test if our current models of moderately fast rotating 
stars are able to describe the observed pulsation modes.

In Fig.~\ref{fig:epscep} we show two amplitude spectra for the \dss\ star \epscepp. 
In the \lee\ panel we used 16 nights obtained from Observatorio 
de Sierra Nevada (OSN) over five months in \str\ $v$ \citep{costa03}
and the \rii\ panel is the result when using two weeks of 
observations with WIRE. Since there are
no significant gaps in the WIRE time series the window function
is much simpler than what is seen in the OSN dataset.
This makes the cleaning of the amplitude spectrum a much easier task.
On the other hand, the much longer time series from OSN makes
the phase determination much more accurate. We are now using
phase differences and amplitude ratios in the \str\ filters
to identify the modes (as was done by \citet{moya04}) seen 
in \dss\ stars observed with WIRE and from OSN \citep{brunttEPSCEP}.

\begin{figure*}[]
\resizebox{\hsize}{!}{\includegraphics[clip=true]{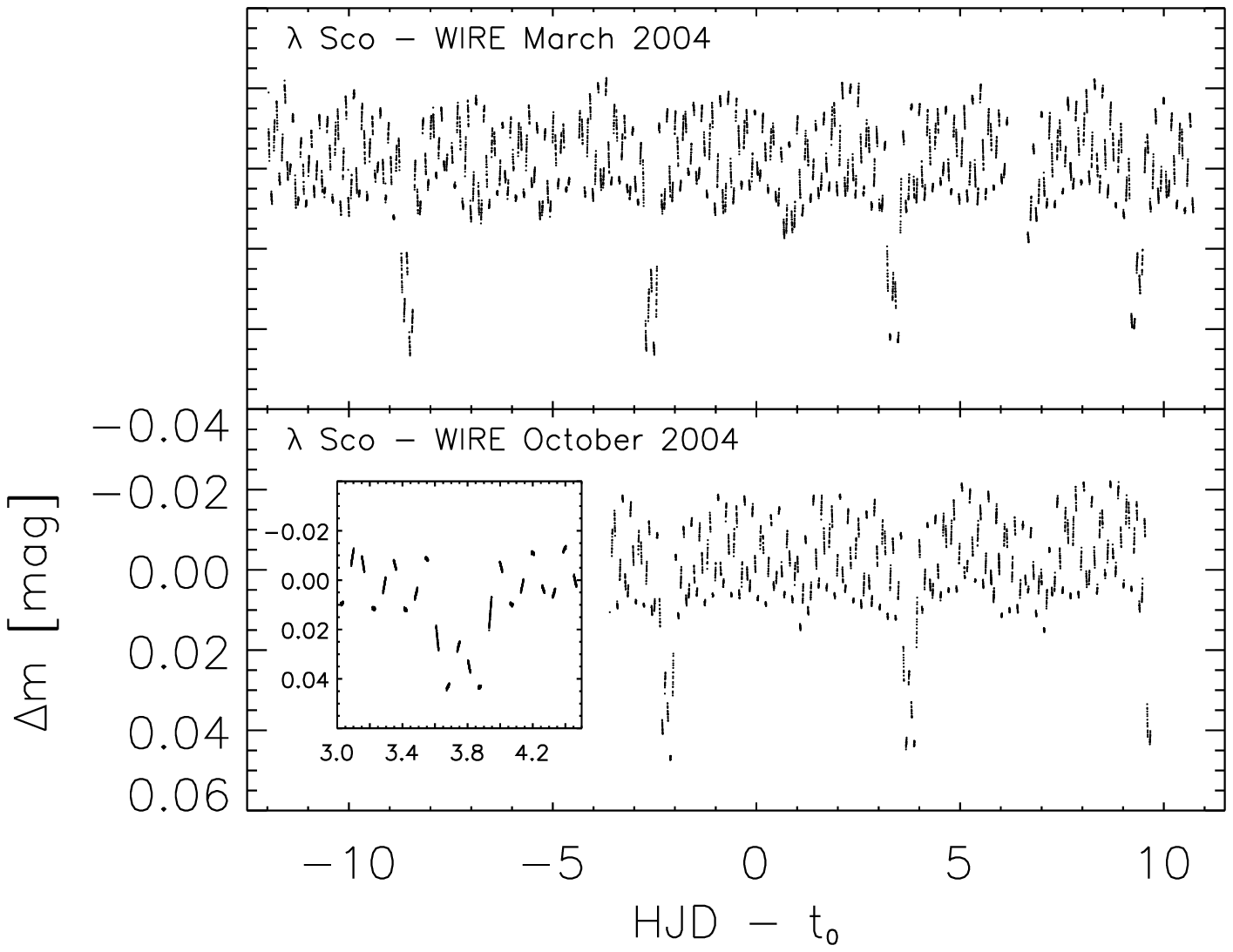}
                      \includegraphics[clip=true]{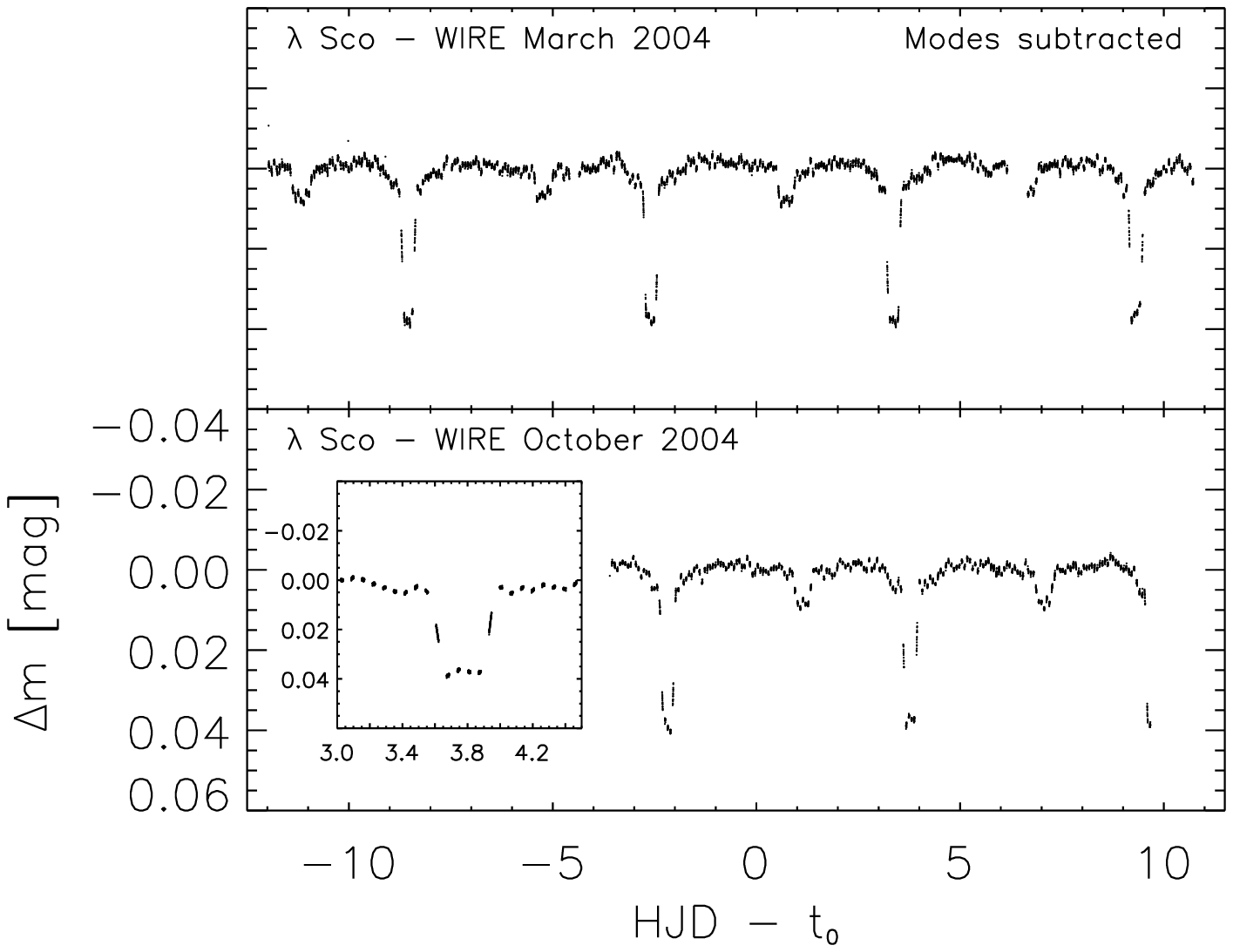}}
\caption{
\footnotesize
Light curves of the \betacep\ star \lamsco\ before and after 
subtracting the intrinsic oscillation modes.
}
\label{fig:lamsco}
\end{figure*}

\section{A \betacepp\ star: \lamsco}

The \lamsco\ system is a triple star system consisting of two
B type stars in a wide orbit ($P\sim1083$\,d). 
One of the B stars has a cooler companion in a close 
orbit \citep{uytterhoeven04} ($P\sim5.95$\,d). 
WIRE observed \lamsco\ twice in 2004 and
the light curves are shown in Fig.~\ref{fig:lamsco}. 
Apart from the main mode of oscillation at 4.680 c/day several 
low amplitude modes are observed. After subtracting the 
intrinsic variation both the primary and secondary eclipses 
are clearly seen. The primary eclipse is barely visible in the 
HIPPARCOS data \citep{uytterhoeven04} but the new data 
allows for a study of the detailed shape and depth of both
the primary and secondary eclipses.

We have modelled the light curve using the 
Wilson-Devinney code \citep{rewd71,rew94}. 
We find that the secondary
star is an A-type star with a mass of 1.8(1)\,$M_\odot$,
radius $R/R_\odot=1.7(1)$, and $T_{\rm eff} = 10\,500\pm500$~K.
This agrees with the preferred
scenario of \citet{uytterhoeven04} and definitely 
rules out a massive white dwarf as suggested by \citet{berghofer00}.
A detailed study of \lamsco\ using the WIRE data 
is in preparation \citep{brunttLAMSCO}.


\section{Future Prospects}

Since the main mission of the WIRE satellite failed 
soon after launch in March 1999 the star tracker has been 
working as a very successful instrument for asteroseismic
studies of a wide range of bright stars. The star
tracker was in use from May 1999 until August 2000
and started again in December 2003. The main goal has been
to study the variability of stars across the HR diagram.
Today we have a database of reduced light curves for
200 stars that have been monitored for typically 2--4 weeks.
In this paper we have presented results for a few 
selected stars observed with WIRE. 

In the coming months we have
coordinated simultaneous observations of a few \dss\ stars 
with WIRE and from the ground with \str\ photometry 
in order to be able to identify the modes.
We plan to conduct two or more WIRE campaigns on some of
the targets in order to achieve a higher frequency 
resolution to be able to resolve closely spaced modes. 
\citet{breger06} notes that the amplitude variation
often seen in \dss\ stars may in fact be due to a combination
of closely spaced modes and inadequate 
frequency resolution; thus an extended time baseline is needed.
Obviously, this applies not only to asteroseismological studies of
\dss\ stars and we have already conducted observations of a
few known \betacep\ stars with six months in between 
(\eg\ \lamsco\ mentioned in this paper) to increase the 
frequency resolution.

\begin{acknowledgements}
HB is grateful for support from the 
Danish FNU and IDA funds.
\end{acknowledgements}

\bibliographystyle{aa}

\end{document}